\documentclass[sigconf]{acmart}

\usepackage{algorithm}
\usepackage{algorithmic}
\usepackage{graphicx}
\usepackage{textcomp}
\usepackage{xcolor}
\usepackage{enumitem}
\usepackage{multirow}
\usepackage{booktabs}
\usepackage{diagbox}
\usepackage{colortbl}
\usepackage{subcaption}
\usepackage{verbatim}
\usepackage{textcomp}
\AtBeginDocument{%
  }

\setcopyright{acmlicensed}
\copyrightyear{2026}
\acmYear{2026}
\setcopyright{cc}
\setcctype{by}
\acmConference[WWW '26]{Proceedings of the ACM Web Conference 2026}{April 13--17, 2026}{Dubai, United Arab Emirates}
\acmBooktitle{Proceedings of the ACM Web Conference 2026 (WWW '26), April 13--17, 2026, Dubai, United Arab Emirates}
\acmPrice{}
\acmDOI{10.1145/3774904.3792808}
\acmISBN{979-8-4007-2307-0/2026/04}

\begin{document}

%%
%% The "title" command has an optional parameter,
%% allowing the author to define a "short title" to be used in page headers.
%\title{SparseCTR: Unleashing the Potential of Long-term Behaviors for Click-Through Rate Predictions}
\title{Unleashing the Potential of Sparse Attention on Long-term Behaviors for CTR Prediction}

%%
%% The "author" command and its associated commands are used to define
%% the authors and their affiliations.
%% Of note is the shared affiliation of the first two authors, and the
%% "authornote" and "authornotemark" commands
%% used to denote shared contribution to the research.
\author{Weijiang Lai}
\affiliation{%
\institution{Institute of Software, Chinese
Academy of Sciences}
\institution{University of Chinese Academy of Sciences
\city{Beijing}
  \country{China}}}
\email{laiweijiang22@otcaix.iscas.ac.cn}

\author{Beihong Jin$^\dagger$}
% \authornote{Corresponding author.}
\affiliation{%
  \institution{Institute of Software, Chinese
Academy of Sciences}
\institution{University of Chinese Academy of Sciences
\city{Beijing}
  \country{China}}}
\email{Beihong@iscas.ac.cn	}

\author{Di Zhang}
\affiliation{%
  \institution{Meituan}
  \city{Beijing}
  \country{China}}
\email{zhangdi77@meituan.com}

\author{Siru Chen}
\affiliation{%
  \institution{Meituan}
  \city{Beijing}
  \country{China}}
\email{chensiru02@meituan.com}

\author{Jiongyan Zhang}
\affiliation{%
  \institution{Meituan}
  \city{Beijing}
  \country{China}}
\email{zhangjiongyan@meituan.com}

\author{Yuhang Gou}
\affiliation{%
  \institution{Meituan}
  \city{Beijing}
  \country{China}}
\email{gouyuhang@meituan.com}

\author{Jian Dong}
\affiliation{%
  \institution{Meituan}
  \city{Beijing}
  \country{China}}
\email{dongjian03@meituan.com}

\author{Xingxing Wang}
\affiliation{%
  \institution{Meituan}
  \city{Beijing}
  \country{China}}
\email{wangxingxing04@meituan.com}

\thanks{$^\dagger$ Corresponding author.}

%%
%% By default, the full list of authors will be used in the page
%% headers. Often, this list is too long, and will overlap
%% other information printed in the page headers. This command allows
%% the author to define a more concise list
%% of authors' names for this purpose.
\renewcommand{\shortauthors}{W. Lai et al.}

%%
%% The abstract is a short summary of the work to be presented in the
%% article.
\begin{abstract}
In recent years, the success of large language models (LLMs) has driven the exploration of scaling laws in recommender systems. However, models that demonstrate scaling laws are actually challenging to deploy in industrial settings for modeling long sequences of user behaviors, due to the high computational complexity of the standard self-attention mechanism. Despite various sparse self-attention mechanisms proposed in other fields, they are not fully suited for recommendation scenarios. This is because user behaviors exhibit personalization and temporal characteristics: different users have distinct behavior patterns, and these patterns change over time, with data from these users differing significantly from data in other fields in terms of distribution. To address these challenges, we propose SparseCTR, an efficient and effective model specifically designed for long-term behaviors of users. To be precise, we first segment behavior sequences into chunks in a personalized manner to avoid separating continuous behaviors and enable parallel processing of sequences. Based on these chunks, we propose a three-branch sparse self-attention mechanism to jointly identify users’ global interests, interest transitions, and short-term interests. Furthermore, we design a composite relative temporal encoding via learnable, head-specific bias coefficients, better capturing sequential and periodic relationships among user behaviors. Extensive experimental results show that SparseCTR not only improves efficiency but also outperforms state-of-the-art methods. More importantly, it exhibits an obvious scaling law phenomenon, maintaining performance improvements across three orders of magnitude in FLOPs. In online A/B testing, SparseCTR increased CTR by 1.72\% and CPM by 1.41\%. Our source code is available at https://github.com/laiweijiang/SparseCTR.

\end{abstract}

\ccsdesc[500]{Information systems~Recommender systems}
\ccsdesc[500]{Information systems~Online advertising}
\ccsdesc[500]{Information systems~Learning to rank}

%%
%% Keywords. The author(s) should pick words that accurately describe
%% the work being presented. Separate the keywords with commas.
% \keywords{Do, Not, Us, This, Code, Put, the, Correct, Terms, for,
%   Your, Paper}
\keywords{CTR Prediction; Sparse Self-attention; Scaling Law}

%% A "teaser" image appears between the author and affiliation
%% information and the body of the document, and typically spans the
%% page.

%%
%% This command processes the author and affiliation and title
%% information and builds the first part of the formatted document.
\maketitle

\section{Introduction}
With the Transformer and LLMs demonstrating impressive performance in various fields~\cite{gpt3,gpt4,scaling_cv1,scaling_modal1,scaling_dense}, the recommender systems community has increasingly explored architectures with self-attention~\cite{llm_rec_survey,gen_review}. Recent studies have shown that these systems not only achieve high performance but also adhere to scaling laws to varying degrees; that is, the system performance will continue to improve as FLOPs (Floating Point Operations) increase~\cite{kaplan2020scaling}. However, the computational complexity of self-attention increases quadratically with the length of the input sequence, hindering efficiency in modeling long-term user behaviors and preventing the system from being deployed online.

%However, the self-attention mechanism, the kernel component of these systems, has obvious deficiencies, since its computational complexity increases quadratically with the input sequence length. When modeling long-term behaviors of users, the self-attention mechanism becomes less efficient, preventing the system from being deployed online.

To deal with this issue, some models resort to randomly sampling user behaviors or focusing only on short-term behaviors to model user interests~\cite{hstu,LSRM}, which inevitably leads to the loss of information and limits the potential of models. Others adopt sparse self-attention~\cite{SUAN}, but these are borrowed from computer vision~\cite{dilated_sparse1,dycoke} or natural language processing~\cite{nsa,dynamic_chunk} fields, which are not well-suited for recommender systems. For instance, dilated or fixed-length chunk sparse self-attention methods~\cite{dilated_sparse1,nsa} typically assume that the input data are uniformly distributed or locally continuous. However, these assumptions do not hold for user behaviors in recommender systems, resulting in the suboptimality of these methods.
%However, due to the randomness and discontinuity of user behaviors, traditional sparse self-attention methods are often suboptimal in this context.
%However, user behaviors in recommender systems are highly non-uniform in time; that is, adjacent behaviors may occur within seconds, or be spread across weeks or even months, making traditional sparse self-attention methods suboptimal in this context.

%We note that in recommendation scenarios, user behaviors often exhibit highly non-uniform temporal characteristics. Similar behaviors within the same period reflect user interests during that period, while substantial variation across different periods indicates shifts in user preferences. Therefore, capturing these dynamics will help compress similar behaviors and identify key behaviors for sparse self-attention. Furthermore, some user behaviors imply sequential and periodic relationships, so efficiently incorporating temporal information into the attention computation can better model dependencies between behaviors.
We note that in recommendation scenarios, user behaviors often exhibit highly personalized and non-uniform temporal characteristics. Differences in behavior distribution exist not only between different users but also within the same user at different times. Specifically, behaviors within the same period reflect user interests during that period, while substantial variation across different periods indicates shifts in user preferences. Therefore, capturing these personalized dynamics will help aggregate redundant behaviors and identify key behaviors for sparse self-attention. Furthermore, given that user behaviors inherently possess sequential and periodic dependencies, explicitly incorporating temporal signals into attention computation enhances the modeling of behavioral interactions.

%Given the highly non-uniform and variable time spans of user behaviors in recommendation scenarios, we observe that temporal discontinuities often signify transitions in user interests. Leveraging this information to design sparse self-attention mechanisms can enable the model to more effectively capture the evolution of user interests. Moreover, user behaviors exhibit rich sequential and periodic patterns, making it essential to efficiently incorporate such temporal information into the attention computation to further capture the dependencies between behaviors.

In this paper, we propose SparseCTR, a model to efficiently and effectively handle long-term behaviors in CTR scenarios. The model primarily consists of stacked SparseBlocks, where EvoAttention (Evolutionary Sparse Self-attention) progressively captures relationships among user behaviors.
EvoAttention first employs a personalized time-aware chunking method named TimeChunking to segment behavior sequences according to time intervals of user behaviors. Based on these chunks, EvoAttention designs global, transition, and local attention mechanisms to model long-term interests, interest transitions, and short-term interests, respectively. 
Additionally, RelTemporal (Relative Temporal Encoding), an efficient relative encoding method, is proposed and incorporated into the attention computation to enable more accurate modeling of complex temporal relationships between behaviors.

Our contributions are summarized as follows:
\begin{itemize}
%\item We propose a personalized method, TimeChunking, for partitioning a user behavior sequence of varying length into a fixed number of chunks. This method maintains high cohesion within each chunk and low coupling between different chunks, and more importantly, enables parallel processing of user sequences. 
%\item We propose a personalized method, TimeChunking, to divide a user behavior sequence into a fixed number of varied-length chunks. This method not only maintains high cohesion within each chunk and low coupling between different chunks but also enables parallel processing of sequences.
\item We propose a personalized method, TimeChunking, to divide different user behavior sequences into the same number of varied-length chunks, maintaining high cohesion within each chunk and low coupling between different chunks, and enabling parallel processing of sequences. 
%\item We propose an efficient method, RelTemporal, to encode the relative time between two behaviors, forming the basis for capturing the sequential and periodic relationships among user behaviors.
\item We propose an efficient method, RelTemporal, to encode the relative time between two behaviors (i.e., time duration, hour, and weekend information) through learnable, head-specific bias coefficients, forming the basis for capturing the sequential and periodic relationships among user behaviors.
\item We present a novel sparse self-attention mechanism, EvoAttention, for chunks of user behaviors with relative time encoding. The mechanism models user interests from multiple perspectives using global, transition, and local attention. 
% \item We propose a novel sparse self-attention mechanism for modeling long-term behaviors of users, which adaptively segments user behavior sequences into chunks based on time intervals and models user interests from multiple perspectives using global, transition, and local attention.
% \item We efficiently incorporate a composite relative temporal encoding into the attention computation to model sequential and periodic relationships among user behaviors.
\item We conduct extensive offline experiments on three real-world datasets and online A/B testing. The results show that SparseCTR achieves state-of-the-art performance while maintaining high efficiency, and exhibits obvious scaling laws; that is, the AUC consistently improves as the FLOPs of SparseCTR increase over three orders of magnitude.
\end{itemize}

\section{Related Work}
\subsection{Long-term Behavior Modeling}
%\subsection{CTR prediction/Long User Behavior Sequences Modeling}
In recent years, CTR prediction has focused on modeling user long-term behaviors to fully capture user interests. Mainstream approaches include one-stage methods~\cite{sdim,SUAN}, which optimize attention for comprehensive modeling, and two-stage methods~\cite{sim,eta,twin,twinv2,diffumin,LONGER}, 
where the newly-added first stage retrieves subsequences for subsequent processing.
%which retrieve subsequences for subsequent processing. 
Inspired by the success of LLMs~\cite{bai2023qwen,llama,deepseek1,deepseek2,deepseek3}, many recent studies extensively explore scaling laws in recommender systems~\cite{clue,Chitlangia2023,scaling_rec}. For instance, 
scaling laws have been studied in some sequential recommendation models (e.g., LSRM)~\cite{LSRM,SRT,actionpiece} and CTR models (e.g., SUAN)~\cite{SUAN,ardalani2022understanding,guo2023embedding,wukong,scaling2025,LONGER}.
%LSRM~\cite{LSRM,SRT,actionpiece} and SUAN~\cite{SUAN,ardalani2022understanding,guo2023embedding,wukong,scaling2025,LONGER} investigate scaling laws in sequential recommendation and CTR prediction, respectively, 
In addition, methods exemplified by HSTU
%while methods like HSTU
~\cite{hstu, onerec1, onerec2,UniROM,onestone} unify retrieval and ranking stages. Despite their promising performance and scalability, deploying these models for long sequences remains challenging due to the high computational cost of self-attention, often necessitating strategies to compress the input or model~\cite{mtgr, SUAN}.

\subsection{Sparse Self-attention Mechanisms}

To address the high time complexity of self-attention~\cite{transformer} when modeling long sequences, various sparse self-attention methods have been proposed~\cite{sparse_survey}. Some methods select a subset of elements for attention computation based on positional or structural information~\cite{star_transformer, etc, bp_transformer, zhao2019explicit}. For example, Longformer~\cite{dilated_sparse1, dilated_sparse2} introduces dilated attention, where each query attends only to keys at fixed intervals. BigBird~\cite{big_bird} proposes random attention by sampling keys randomly. Recent 
%fixed-length chunk attention 
methods such as NSA and MoBA~\cite{moba, nsa} divide sequences into fixed-length chunks, where each query first attends to chunk representations and then interacts with elements in the top-$k$ chunks. However, these methods often assume uniform or locally continuous data distributions, which differ from the highly non-uniform user behavior characteristics observed in recommendation scenarios. Besides, clustering attention~\cite{clusterformer1, clusterformer2,routing} allows each query to attend only to similar elements that are identified by clustering techniques. A representative case is Reformer~\cite{reformer}, which employs hash encoding~\cite{simhash} to group elements in a sequence. However, these methods disrupt the original sequence order and are not suitable for order-sensitive scenarios.

Another line of research forms kernelized attention. Some methods~\cite{Linformer, performer, Synthesizer}, such as Linformer and Performer, approximate attention by projecting $Q$ and $K$ into lower-dimensional spaces using kernel functions or matrix decomposition. Some methods~\cite{Transformers_are_rnns, linrec, Efficient_Attention, luna}, such as Nyströmformer, reduce complexity by using kernel functions $\phi$ to approximate $\text{softmax}(QK^\top)V$ as $\phi(Q)(\phi(K)^\top V)$, thus avoiding the quadratic-order computational cost. However, these methods are in essence approximations of full attention and are difficult to extend with relative information encoding.
%the $O(n^2)$ cost. However, these methods are still approximations of full attention and are difficult to extend with relative information encoding.

%Another line of work, known as clustering attention and represented by Reformer~\cite{reformer, clusterformer1, clusterformer2,routing}, employs clustering techniques such as hashing encoding~\cite{simhash} to group sequence elements, enabling each query to attend only to similar elements. 
%Another line of research, termed clustering attention~\cite{reformer, clusterformer1, clusterformer2,routing}, allows each query to attend only to similar elements that are identified by clustering techniques. A representative case is Reformer~\cite{reformer}, which employs hash encoding~\cite{simhash} to group elements in a sequence. 
%However, these methods disrupt the original sequence order and are not suitable for order-sensitive scenarios.

\subsection{Relative Information Encoding}
Relative position encoding is a key technique in attention mechanisms~\cite{position_survey}. Unlike absolute position encoding~\cite{transformer,absolute_position}, it captures the relative distances between elements in a sequence, thereby improving model performance. Shaw et al.~\cite{shaw2018self} first introduce learnable relative position embeddings for each query-key pair, which are combined with the $K$ or $V$ matrices. This idea has been further extended in multiple systems such as Music Transformer~\cite{huang2018music, hstu}, T5~\cite{t5}, and Transformer-XL~\cite{Transformer_XL}, each of which adopts learnable relative encoding and achieves excellent results. Other methods, including DeBERTa~\cite{he2020deberta, tupe}, propose disentangled encoding, where the content and position of an element are encoded separately and their attention scores are combined into one. While effective, these methods incorporate additional lookup operations or computational overhead into the attention mechanism, which potentially decreases their efficiency.

Recently, some methods such as Alibi~\cite{alibi,chi2022kerple,Sandwich} adopt a simpler strategy by directly adding a negative bias proportional to the relative position in the attention scores, thereby greatly improving efficiency. Sinusoidal methods such as RoPE~\cite{rope, longrope} encode relative position by rotating queries and keys, allowing their dot product to capture relative positional information. Note that the efficient integration of temporal information into attention computation remains underexplored.

\section{Methodology}

\begin{figure*}[tb]
  \includegraphics[width=\textwidth]{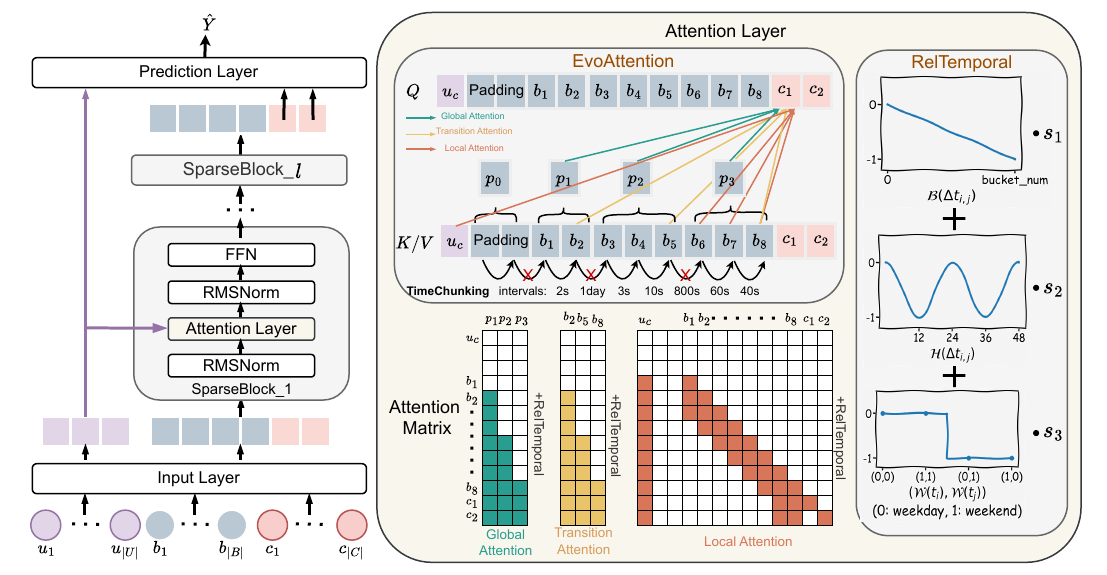}
  \caption{Architecture of SparseCTR.}
  \label{fig:model}
\end{figure*}
\subsection{Problem Formulation}
For a user, his/her historical behaviors are denoted as $B = \{b_i\}_{i=1}^{|B|}$, where each behavior $b_i$ includes features such as item ID and behavior time. The features of user profile are represented as $U = \{u_i\}_{i=1}^{|U|}$, such as user age. Here, $|B|$ and $|U|$ denote the number of behaviors and user profile features, respectively. For a target candidate $c$, more detailed features are available, such as its CTR over the last 3 days.

Given a user with profile features $U$ and historical behaviors $B$, a set of target candidates $C = \{c_i\}_{i=1}^{|C|}$ and their interaction labels $Y=\{y_i\}_{i=1}^{|C|}$, where $|C|$ is the number of candidates. For each candidate $c_i$, with its label $y_i \in \{0, 1\}$, our task is to predict its click-through probability, which can be formalized as follows:
\begin{equation}
    \mathcal{P}(y_i=1| x_i \in (U, B, c_i))=F(x_i; \theta), \quad \forall c_i \in C, \quad \forall y_i \in Y,
\end{equation}
where $F(x_i; \theta)$ is the model to be developed and $\theta$ represents the model parameters.

For this task, we propose a model named SparseCTR. Figure~\ref{fig:model} shows the architecture of our model, which includes the input layer, multiple SparseBlocks, and the prediction layer.

\subsection{Input Layer}
% In the decoder-only architecture, each element is updated based solely on its preceding elements. Leveraging this property, 
%We concatenate multiple target candidates after the user behavior sequence and model them jointly, using a masking mechanism that ensures each element in the sequence can only attend to previous elements and that targets are isolated from each other. This enables the model to predict all targets in a single inference process, greatly improving efficiency compared to traditional CTR frameworks that predict each target individually. To ensure consistency between offline and online settings, we aggregate a user's pairwise samples within a single exposure, i.e., $(U, c_1, y_1), (U, c_2, y_2), \ldots, (U, c_{|C|}, y_{|C|})$ into a listwise sample $(U, C, Y)$, where $C = \{c_1, ..., c_{|C|}\}$ and $Y = \{y_1, ..., y_{|C|}\}$.
For a user, online CTR prediction requires ranking multiple candidates. Traditional CTR frameworks predict each candidate individually, resulting in redundant computation and low efficiency. 
To address this, we make those SparseBlocks work in the manner of a causal encoder. To this end, we concatenate candidates behind the user behavior sequence and apply a masking mechanism that ensures that each element only attends to previous elements and that target candidates remain isolated.
%To address this, we utilize a causal encoder to jointly model multiple target candidates by concatenating them after the user behavior sequence and applying a masking mechanism that ensures each element only attends to previous elements and that target candidates remain isolated. 
This enables all candidates to be predicted in a single inference, greatly improving efficiency. 

To maintain consistency between offline training and online inference, we aggregate a user's pairwise samples within a single exposure when constructing offline training data, i.e., $\{(U, B, c_1, y_1), \\ \ldots, (U, B, c_{|C|}, y_{|C|})\}$, into a listwise sample $(U, B, C, Y)$, where $C = \{c_1, \ldots, c_{|C|}\}$ and $Y = \{y_1, \ldots, y_{|C|}\}$. 
Further, in the input layer, we construct a mixed sequence $S = \{b_1, \ldots, b_{|B|}, c_1, \ldots, c_{|C|}\}$, and use a uniform embedding table to initialize its embedding matrix $E_S \in \mathbb{R}^{n \times d}$ and the embedding vector of user features $e_U \in \mathbb{R}^{|U|d}$, where $n = |B| + |C|$. The features of user behavior are concatenated and mapped to the embedding with dimension $d$, and the features of candidates undergo the same process.
%The features of user behaviors and candidates are concatenated separately and mapped to the same embedding dimension $d$.
%In the input layer, we concatenate user behaviors and target candidates to form a mixed sequence $S = \{b_1, \ldots, b_{|B|}, c_1, \ldots, c_{|C|}\}$, and use a uniform embedding table to obtain its embedding matrix $E_S \in \mathbb{R}^{n \times d}$ and the embedding vector of user features $e_U \in \mathbb{R}^{|U|d}$, where $n = |B| + |C|$. The features of user behaviors and targets are concatenated separately and mapped to the same embedding dimension $d$.

\subsection{SparseBlock}
To capture complex relationships among user behaviors, we stack $l$ SparseBlocks to extract higher-level and more abstract behavioral features progressively. Each SparseBlock consists of RMSNorm, an attention layer, and a Feedforward Network (FFN).

In SparseBlock, we first apply RMSNorm~\cite{rmsnorm} to $E_S$ for pre-normalization, which helps stabilize training. RMSNorm, widely adopted in large language models~\cite{llama,bai2023qwen}, normalizes the input based on its root mean square. Compared to standard LayerNorm, it reduces computational cost while maintaining training stability.

The normalized embeddings are then passed to the attention layer, where EvoAttention captures global, transition, and local dependencies among user behaviors and RelTemporal incorporates the temporal information into the attention computation,
%which efficiently captures global, transition, and local dependencies among user behaviors. Furthermore, temporal information is incorporated into the attention computation via RelTemporal, 
enabling the model to capture the evolution of user interests and fine-grained temporal relationships in user behavior sequences.

Finally, the output of EvoAttention $E_s^{Evo}$ is fed into a SwiGLU-based FFN to enhance the model's nonlinear transformation and representation capacity. The computation is defined as follows:
\begin{equation}
    E_{S}^{FFN}=(\phi (E_{S}^{Evo}W_1) \odot E_{S}^{Evo}W_2)W_3,
\end{equation}
where $\phi$ denotes the Swish activation function \cite{swish}, and $W_1, W_2 \in \mathbb{R}^{d \times 3d}, W_3 \in \mathbb{R}^{3d \times d}$ are learnable parameters. This structure combines the smooth, differentiable properties of Swish with the gating mechanism of GLU~\cite{glu}, enabling the model to effectively capture complex feature interactions and enhance its expressive power.

In the following, we provide detailed descriptions of EvoAttention and RelTemporal, which are essential for modeling multi-granularity dependencies and temporal relationships among user behaviors.

\subsection{Evolutionary Sparse Self-attention}
Self-attention mechanisms are widely used to model complex relationships among elements in a sequence. However, their quadratic complexity limits efficiency when handling long sequences. Although various sparse self-attention methods have been proposed, the unique distribution of user behaviors in recommender systems renders these methods less suitable for recommendation tasks.

To address this issue, we propose EvoAttention, an efficient method designed based on the distributional characteristics of user behaviors to model key dependencies in behavior sequences. This method follows our finding that user behaviors exhibit personalization and temporal characteristics: different users have distinct patterns that evolve. For each user, behaviors within the same period tend to be similar, whereas across different periods, the number and distribution of behaviors can change significantly.

%user behaviors within the same period tend to be more similar, while the number and distribution of behaviors can vary significantly across different periods.

\subsubsection{Personalized Time-aware Chunking}
To capture the continuous behaviors and evolving interests of users, we propose TimeChunking, a personalized time-aware method that segments different behaviors into the same number of varied-length chunks. This approach compresses sequences based on personalization and temporal characteristics while enabling parallel processing for the self-attention mechanism.
%To effectively capture the continuous behaviors and evolving interests of users, we propose a personalized, time-aware chunking method called TimeChunking, which integrates the behaviors of different users into a fixed number of varied-length chunks, compressing sequences based on personalization and temporal characteristics while ensuring that the model can process sequences in parallel.
%To effectively capture individual users' continuous behaviors and the evolution of their interests, we introduce a personalized, time-aware chunking method called TimeChunking.
%To effectively capture continuous behaviors and the evolution of user interests, we introduce a personalized time-aware chunking method TimeChunking. 

Specifically, we calculate the time intervals between adjacent behaviors and select the top-$|P|$ largest intervals as segmentation points. Behaviors between two points form a chunk, and since padding behaviors have a time value of 0, the interval between the last padding and the first valid behavior is maximized, naturally forming a separate padding chunk $p_0$. This results in a set of valid behavior chunks $P = \{p_1, ..., p_{|P|}\}$.
%Specifically, we calculate the time intervals between adjacent behaviors and select the top-$|P|$ largest intervals as segmentation points. Behaviors between two segmentation points are grouped into a chunk, while padding behaviors with a designated padding time value of 0 naturally form a separate padding chunk $p_0$. This results in a set of valid behavior chunks $P = \{p_1, ..., p_{|P|}\}$.

% Based on behavior chunks, we further design global, transition, and local attention mechanisms to model user global interests, interest transitions, and short-term interests, respectively, thereby enabling multi-granular modeling of user behaviors. In the implementation, we first apply linear transformations to the sequences:

Building on these chunks, we design global, transition, and local attention mechanisms to model user behaviors from multiple perspectives. In the implementation, we first apply linear transformations to the sequences:
\begin{equation}
    Q = E_S W_Q,\quad K = E_B W_K,\quad V = E_B W_V,
\end{equation}
where $E_S$ and $E_B$ denote the embedding matrices of the full mixed sequence and the behavior part, respectively, and $W_Q, W_K, W_V \in \mathbb{R}^{d \times d}$ are learnable parameters.

\subsubsection{Global Attention}
% \noindent\textbf{Global Attention.}
To enhance the model's capability to capture global dependencies, we introduce a global attention mechanism to model users' long-term interests. Specifically, we aggregate the key and value vectors within each behavior chunk to obtain chunk-level key and value representations, as follows:
\begin{equation}
    k_{p_i} = \text{Aggregate}(\{k_{b_j} \mid b_j \in p_i\}), \quad \forall p_i \in P,
\end{equation}
\begin{equation}
    v_{p_i} = \text{Aggregate}(\{v_{b_j} \mid b_j \in p_i\}), \quad \forall p_i \in P,
\end{equation}
where $\{k_{b_j} \mid b_j \in p_i\}$ and $\{v_{b_j} \mid b_j \in p_i\}$ are vectors from $K$ and $V$ associated with behaviors in the $i$-th chunk, and $\text{Aggregate}(\cdot)$ is the aggregation operation, which is implemented as a multi-layer perceptron (MLP).

Next, we model the relationships between each element and all preceding behavior chunks, computed as follows:
\begin{equation}
    \text{Attention}(Q, K_P, V_P) = \text{softmax}\left(\frac{QK_P^\top}{\sqrt{d}}\right)V_P,
\label{attention1}
\end{equation}
where $K_P = [k_{p_1}, \ldots, k_{p_{|P|}}]$ and $V_P = [v_{p_1}, \ldots, v_{p_{|P|}}]$ are the chunk-level key and value embedding matrices, with $K_P, V_P \in \mathbb{R}^{|P| \times d}$.

\subsubsection{Transition Attention}
In recommendation scenarios, the last few behaviors in a user's consecutive behavior chunk often indicate the user's current interests or potential shifts in interest. Effectively capturing such transition behaviors is crucial for accurately modeling user preferences and improving the prediction of future actions. Therefore, we select the last $m$ behaviors from each valid chunk as transition behaviors, forming the set $B' = \{b'_1, ..., b'_{m|P|}\}$, and model the dependencies between each element and all preceding transition behaviors as follows:
\begin{equation}
    \text{Attention}(Q, K_{B'}, V_{B'}) = \text{softmax}\left(\frac{QK_{B'}^\top}{\sqrt{d}}\right)V_{B'},
\label{attention2}
\end{equation}
where $K_{B'} = [k_{b'_1}, \ldots, k_{b'_{m|P|}}]$ and $V_{B'} = [v_{b'_1}, \ldots, v_{b'_{m|P|}}]$ are the key and value embedding matrices for the transition behaviors, i.e., $k_{b'_i} \in K$, $v_{b'_i}\in V$, and $K_{B'}, V_{B'} \in \mathbb{R}^{m|P| \times d}$.

\subsubsection{Local Attention}
To better capture user interests within short time windows while accounting for the influence of user profile features on sequence elements, we design a local attention mechanism. For each element $i$, its local behavior set is defined as the preceding $w$ behaviors, $B''_i = \{u_c, b_{i-(w-1)}, \ldots, b_i\}$, where $u_c$ is a representation of $|U|$ user features compressed via an MLP. We then model the relationships between each element and its local behaviors as follows:
% To better capture user interests within short time windows, we design a local attention mechanism. For each element $i$, we define its local behavior set as the preceding $w$ behaviors, $B''_i = \{b_{i-(w-1)}, \ldots, b_i\}$. We then model the relationships between each element and its local behaviors as follows:
\begin{equation}
    \text{Attention}(Q, K_{B''}, V_{B''}) = \text{softmax}\left(\frac{QK_{B''}^\top}{\sqrt{d}}\right)V_{B''},
\label{attention3}
\end{equation}
where $K_{B''} = \{K_{B''_i}\}_{i=1}^n$ and $V_{B''} = \{V_{B''_i}\}_{i=1}^n$ are the key and value embeddings for all windows, respectively. For the $i$-th query, $K_{B''_i} = [k_{b_{i-(w-1)}}, \ldots, k_{b_{i}}]$ and $V_{B''_i} = [v_{b_{i-(w-1)}}, \ldots, v_{b_{i}}]$ are the key and value matrices for the local window, i.e., $k_{b_{i}} \in K$, $v_{b_{i}} \in V$, and $K_{B''_i}, V_{B''_i} \in \mathbb{R}^{w \times d}$.

\subsubsection{Sparse Self-attention Fusion}
To integrate global, transition, and local attention and fully exploit multi-granularity interest information, we fuse the outputs of all three branches using a gating mechanism that adaptively learns the importance of each branch. The fusion process is as follows:
% \begin{align}
% [\alpha_1, \alpha_2, \alpha_3] =\ &\text{softmax}\Big( [\text{Attention}(Q, K_P, V_P),\ \text{Attention}(Q, K_{B'}, V_{B'}), \notag\\
% &\quad \text{Attention}(Q, K_{B''}, V_{B''})] W_{\text{gate}} \Big)
% \end{align}
\begin{multline}
[\alpha_1, \alpha_2, \alpha_3] = \text{softmax} \Big( [\text{Attention}(Q, K_P, V_P), \\
\text{Attention}(Q, K_{B'}, V_{B'}),\ \text{Attention}(Q, K_{B''}, V_{B''})] W_{\text{gate}} \Big),
\end{multline}
\begin{multline}
\text{Attention}(Q,K,V) = \alpha_1 \cdot \text{Attention}(Q, K_P, V_P) \\
+ \alpha_2 \cdot \text{Attention}(Q, K_{B'}, V_{B'}) 
+ \alpha_3 \cdot \text{Attention}(Q, K_{B''}, V_{B''}),
\end{multline}
where $\alpha_1, \alpha_2, \alpha_3$ are the gating weights for the three attention branches, and $W_{\text{gate}} \in \mathbb{R}^{3d \times 3}$ is a learnable matrix.

In practice, we apply multi-head attention in each branch, and the final output is computed as:
\begin{equation}
    E_S^{Evo} = [\text{Attention}_{1}(Q,K,V), ..., \text{Attention}_{H}(Q,K,V)] W_O,
\end{equation}
where $\text{Attention}_i(Q,K,V)$ denotes the $i$-th attention head, $H$ is the number of heads, and $W_O \in \mathbb{R}^{d \times d}$ is a learnable projection matrix. By combining multi-head attention and the gated fusion mechanism, the model adaptively captures dependencies in different subspaces and branches, significantly enhancing the overall modeling capability.

\subsection{Relative Temporal Encoding}
Positional information is crucial for capturing dependencies in attention computation. While absolute positional encoding is commonly used in Transformers, existing studies~\cite{shaw2018self, position_survey} have shown that modeling relative positions can better capture dependencies among elements and improve performance. However, in recommender systems, relative position alone is often insufficient. For instance, behaviors that are distant in the sequence may actually occur close together in time. Moreover, user behaviors often exhibit clear periodic patterns, when examined by differences between times of day, weekdays, or weekends.
%such as differences between times of day or between weekdays and weekends.
%Positional information is crucial for capturing dependencies in attention computation. Traditional Transformers typically use absolute positional encoding, assigning a unique position vector to each element in the sequence. However, subsequent studies~\cite{shaw2018self,position_survey} have shown that modeling relative positional information can more effectively capture dependencies between elements and further improve model performance.

 To more accurately model the relationships between behaviors, we propose RelTemporal, which efficiently incorporates relative time, hour, and weekend information into the attention computation. This enables the model to capture richer and finer-grained temporal dependencies.

\subsubsection{Relative Time Encoding}
Since the range of relative time differences is large and unevenly distributed, we use a bucketing strategy to discretize time intervals when computing pairwise differences. Specifically, for any two behaviors $b_i$ and $b_j$ occurring at times $t_i$ and $t_j$, the time interval is defined as $\Delta t_{i,j} = |t_i - t_j|$. The relative time encoding is then computed as:
% \begin{equation}
%     \mathcal{B}(\Delta t_{i,j}) = \left\lfloor \log_2(\Delta t_{i,j}) \right\rfloor,
% \end{equation}
\begin{equation}
    \text{bias1}_{i,j}^{(h)} = -\mathcal{B}(\Delta t_{i,j}) \cdot s_1^{(h)},\quad h \in \{1, \ldots, H \},
\end{equation}
where $\mathcal{B}(\cdot)= \left\lfloor \log_2(\Delta t_{i,j})\right\rfloor$ is the bucketing function, and $s_1^{(h)}$ is the learnable bias coefficient for the $h$-th attention head~\cite{alibi}. The initial value of $s_1^{(h)}$ is set as a geometric sequence across attention heads; specifically, the $h$-th head is initialized with $\left(2^{\frac{-8}{H}}\right)^{h-1}$. This initialization provides each head with a different sensitivity to time information, enabling more robust modeling of dynamic relationships between elements through temporal information.

\subsubsection{Relative Hour Encoding}
User behaviors often exhibit daily periodicity, with distinct trends at different hours of the day. To capture this phenomenon, we introduce a relative hour encoding to model periodic relationships at the hourly level. Given the cyclic nature of hours (e.g., the hour difference between 0 and 23 o'clock, or between 0 and 1 o'clock, is both 1 hour), we employ a sine function to model this periodicity. The computation is as follows:
\begin{equation}
    \text{bias2}_{i,j}^{(h)} = -\sin\left(\pi \cdot \frac{\mathcal{H}(\Delta t_{i,j})}{24}\right) \cdot s_2^{(h)},\quad h \in \{1, \ldots, H \},
\end{equation}
where $\mathcal{H}(\cdot)$ maps the time interval to the hour difference, and $s_2^{(h)}$ is a learnable parameter initialized in the same way as $s_1^{(h)}$.

\subsubsection{Relative Weekend Encoding}
In addition, user behaviors also show clear periodicity between weekdays and weekends. To model this relationship, we introduce a piecewise function:
\begin{equation}
    \text{bias3}_{i,j}^{(h)} = 
\begin{cases}
  0, & \mathcal{W}(t_i) = \mathcal{W}(t_j) \\
  -1, & \mathcal{W}(t_i) \neq \mathcal{W}(t_j)
\end{cases} \cdot s_3^{(h)},\quad h \in \{1, \ldots, H\},
\end{equation}
where $\mathcal{W}(\cdot)$ is an indicator function that returns $1$ if the time corresponds to a weekend and $0$ otherwise, and $s_3^{(h)}$ is a learnable parameter initialized in the same way as $s_1^{(h)}$ and $s_2^{(h)}$.

\subsubsection{Relative Temporal Integration}
We assign distinct bias coefficients as learnable weights to each type of temporal information, allowing them to adaptively adjust their contributions when modeling relationships between elements. The complete RelTemporal formula is as follows:
\begin{equation}
\text{bias}_{i,j}^{(h)} = \text{bias1}_{i,j}^{(h)} + \text{bias2}_{i,j}^{(h)} + \text{bias3}_{i,j}^{(h)}, \quad h \in \{1, \ldots, H\},
\end{equation}
In our implementation, the complete RelTemporal is added to each attention branch, resulting in the following modifications to the attention computation in Eqs.~\ref{attention1}, \ref{attention2}, and \ref{attention3}:
\begin{equation}
\text{Attention}(Q, K, V) = \text{softmax}\left(\frac{QK^\top}{\sqrt{d}} + \text{bias}\right) V,
\end{equation}
where $K$ and $V$ denote the key and value embedding matrices for each attention branch, and $\text{bias}$ is the corresponding bias term added to each query-key pair during the attention calculation. For global attention, RelTemporal is computed based on the mean time of each chunk.

% where $K_*,V_*$ and $\text{bias}$ refers to the complete $\text{bias}_{i,j}^{(h)}$ used in the attention calculation. For global attention, the RelTemporal is computed based on the mean timestamp of each chunk.

\subsection{Prediction Layer}
In the prediction layer, we extract the embeddings of the target candidates $E_C^{(l)}$ from the mixed sequence output of the $l$-th SparseBlock, fuse the user feature vector $e_U$ with the candidate representation and feed the fused result into an MLP for CTR prediction.
% $E_S^{(l)}$：这个变量有定义但未使用，故删除
%In the prediction layer, we obtain the mixed sequence output from the $l$-th SparseBlock, denoted as $E_S^{(l)}$, and extract the embeddings of the target candidates $E_C^{(l)}$. To better model the interaction between the user and candidates, we fuse the user feature vector with the candidate representation and feed the fused result into an MLP for CTR prediction. 
The computations are as follows:
\begin{equation}
    \hat{Y} = \text{MLP}(\text{ReLU}(E_C^{(l)} W_4) \odot \text{sigmoid}(e_U W_5)),
\end{equation}
% \begin{equation}
%     E_C^{(l)'} = \text{ReLU}(E_C^{(l)} W_4),
% \end{equation}
% \begin{equation}
%     e_U' = \text{sigmoid}(e_U W_5),
% \end{equation}
% \begin{equation}
%     \hat{Y} = \text{MLP}(E_C^{(l)'} \odot e_U'),
% \end{equation}
where $W_4 \in \mathbb{R}^{d \times d}$ and $W_5 \in \mathbb{R}^{|U|d \times d}$ are learnable parameter matrices. The model is optimized using the binary cross-entropy loss over all target candidates:
\begin{equation}
    \mathcal{L} = -\frac{1}{N |C|} \sum_{i=1}^N \sum_{j=1}^{|C|} \left[ y_{ij} \log \hat{y}_{ij} + (1 - y_{ij}) \log (1 - \hat{y}_{ij}) \right],
\end{equation}
where $N$ is the number of samples in a batch, 
$|C|$ is the number of target candidates per sample. 
$y_{ij}$ and $\hat{y}_{ij}$ denote the ground-truth label and the predicted value for the $j$-th candidates in the $i$-th sample, respectively.

\subsection{Complexity Analysis}
\subsubsection{Time Complexity}
The time complexity of SparseCTR is primarily composed of the stacked $l$ SparseBlocks. The main computational cost in SparseBlocks comes from EvoAttention, whose time complexity is $O \left(\mathcal{B}l(n|P|d + nm|P|d + nwd)\right)$, where $\mathcal{B}$ is the batch size. Since $|P|,m,w \ll n$, the overall complexity is significantly lower than that of full attention, which is $O (\mathcal{B}ln^2d)$.
% 红色B：变量重定义了，在line 254，B 表示 用户行为序列。
\subsubsection{Space Complexity}
In addition to the embedding table, the learnable parameters in SparseCTR mainly come from the SparseBlocks. Specifically, the parameters are primarily from EvoAttention and FFN, with complexities of $O(4l d^2 + 3l d)$ and $O(9l d^2)$, respectively. Compared to full attention, our model introduces fewer additional parameters.

\section{Experiments}
% In this section, we conduct extensive experiments to answer the following Research Questions (RQs):
% \begin{itemize}
% \item \textbf{RQ1}: Does SparseCTR outperform existing CTR models?
% \item \textbf{RQ2}: How do the different branches of EvoAttention and the various components of RelTemporal affect AUC?
% \item \textbf{RQ3}: How do EvoAttention and RelTemporal compare with similar methods?
% \item \textbf{RQ4}: How much can EvoAttention improve efficiency and effectiveness over full self-attention under different configurations?
% \item \textbf{RQ5}: Does SparseCTR holds on scaling laws?
% \item \textbf{RQ6}: How does SparseCTR perform in a live production environment?
% \end{itemize}

\subsection{Experimental Settings}
\subsubsection{Datasets}
We select one industrial dataset and two public datasets to conduct experiments. The statistics of three datasets are shown in Table~\ref{tab:datsets}.

\begin{itemize}
\item \textbf{Industry}: This dataset involves over 86 million users who were active in 7 days prior to behavior recording, containing the historical behaviors of these users over the past 2 years.
%This is an industrial dataset comprising over 86 million users who have been active in the past 7 days. It records their complete behavioral histories spanning the last two years.
%, with each user's behavior annotated with features such as item ID, behavior type, and more. 
\item \textbf{Alibaba}\footnote{https://tianchi.aliyun.com/dataset/56}: 
This dataset, released by Alibaba, collects user behavior data of 22 days from its display advertising system.
%is collected from their display advertising system.
%It contains behavior data from users over a span of 22 days, providing comprehensive information on users, advertisements, and user behaviors.
\item \textbf{Ele.me}\footnote{https://tianchi.aliyun.com/dataset/131047}: 
This dataset is derived from the click logs of the ele.me service, containing user behaviors over 30 days.
%This dataset is constructed from click logs of the ele.me service and records user behaviors over a 30-day period. It includes features related to users, candidate items, and user behaviors.
\end{itemize}

\subsubsection{Competitors}
We select ten representative models from different research lines as baselines. Two of them, i.e., \textbf{DIN}~\cite{din} and \textbf{CAN}~\cite{can}, focus on modeling behavior feature interactions (assigned to Group \uppercase\expandafter{\romannumeral1}). The other eight methods, i.e., \textbf{SoftSIM}, \textbf{HardSIM}, \textbf{ETA}~\cite{eta}, \textbf{TWIN-V2}~\cite{twin}, \textbf{BST}~\cite{bst}, \textbf{HSTU}~\cite{hstu}, \textbf{LONGER}~\cite{LONGER}, and \textbf{SUAN}~\cite{SUAN}, adopt behavior sequence modeling, where \textbf{SoftSIM} and \textbf{HardSIM} are different variants of \textbf{SIM}~\cite{sim}, employing embedding similarity retrieval and category retrieval, respectively. Of these eight methods, the first four are designed for long sequences (assigned to Group \uppercase\expandafter{\romannumeral2}), while the last four are not (Group \uppercase\expandafter{\romannumeral3}).

\begin{table}[t]
\centering
\caption{Statistics of the datasets.}
\setlength\tabcolsep{4.2pt}
\begin{tabular}{ccccccc}
\hline
Datasets        & \#Users & \#Items & \#Samples \\
\hline
Industry & 86,722,000   & 20,241,000  & 1,455,722,000   \\
Alibaba  & 1,141,729   &  461,527  & 700,000,000         \\
Ele.me  & 14,427,689   & 7,446,116   & 128,000,000         \\
\hline
\end{tabular}
\label{tab:datsets}
\end{table}

\begin{table*}[tbp]
  \caption{Performance comparison. For each row, the highest and second-highest AUCs are denoted by bold and underlined text, respectively. The DIN model serves as the base model for computing RelaImpr.}
   \label{tab:performance_comparison}
  \centering
\begin{tabular}{cc|cc|cccc|cccc|c}
    \toprule
    \multirow{2}{*}{Dataset} & \multirow{2}{*}{Metric} & \multicolumn{2}{c|}{Group \uppercase\expandafter{\romannumeral1}} & \multicolumn{4}{c|}{Group \uppercase\expandafter{\romannumeral2}} & \multicolumn{4}{c|}{Group \uppercase\expandafter{\romannumeral3}} \\
    & & DIN & CAN & SoftSIM & HardSIM & ETA & TWIN-V2 & BST & HSTU & LONGER & SUAN & SparseCTR\\
    \midrule
    \multirow{2}{*}{Industry} & AUC & 0.6920 & 0.6918 & 0.6936 & 0.6931 & 0.6944 & 0.6948 & 0.6969 & 0.7002 & 0.7007 & \underline{0.7040} & \textbf{0.7083}$\pm$0.00004\\
    & RelaImpr &0.00\% & -0.10\% & 0.83\% & 0.57\% & 1.25\% & 1.45\% & 2.55\% & 4.27\% & 4.53\% & \underline{6.25\%} & \textbf{8.49\%}$\pm$0.02\% \\
    \midrule
    \multirow{2}{*}{Alibaba} & AUC & 0.6209 & 0.6198 & 0.6227 & 0.6247 & 0.6231 & 0.6237 & 0.6422 & 0.6448 & 0.6457 & \underline{0.6479} & \textbf{0.6530}$\pm$0.00014 \\
    & RelaImpr & 0.00\% & -0.91\% & 1.49\% & 3.14\% & 1.82\% & 2.32\% & 17.62\% & 19.77\% & 20.51\% & \underline{22.33\%} & \textbf{26.55\%}$\pm$0.12\% \\
    \midrule
    \multirow{2}{*}{Ele.me} & AUC & 0.6378 & 0.6382 & 0.6414 & 0.6405 & 0.6410 & 0.6423 & 0.6616 & 0.6645 & 0.6650 & \underline{0.6673} & \textbf{0.6725}$\pm$0.00025 \\
    & RelaImpr & 0.00\% & 0.29\% & 2.61\% & 1.96\% & 2.32\% & 3.27\% & 17.28\% & 19.38\% & 19.74\% & \underline{21.41\%} & \textbf{25.18\%}$\pm$0.18\% \\
    \bottomrule
\end{tabular}
\end{table*}

% Our competitors include the classical CTR models and the CTR models for long-term user behaviors.
% \textbf{DIN}\cite{din}: employs the target attention to dynamically reweight users’ historical behaviors w.r.t. the candidate.
% \textbf{CAN}\cite{can}: disentangles representation learning and feature interaction modeling via a co-action unit
% \textbf{SIM} \cite{sim}: proposes to model the long-term sequences by the general search unit and exact search unit. The variants using the embedding similarity retrieval and category retrieval are called SoftSIM and HardSIM, respectively.
% \textbf{ETA} \cite{eta}: uses SimHash to replace the embedding similarity retrieval approach provided by SoftSIM, which greatly accelerates the retrieval speed.
% \textbf{TWIN-v2} \cite{twin}: adopts the two-stage framework with a new target attention network, and classifies behavioral features into inherent and user-item cross features to efficiently calculate behavior importance scores.

% \textbf{BST} \cite{bst}:
% \textbf{HSTU} \cite{hstu}: 
\begin{table}[t]
\centering
\caption{Results of the ablation study.}
\begin{tabular}{c|ccc}
\hline
Model  & Industry & Alibaba & Ele.me \\
\hline
SparseCTR    & \textbf{0.7083}  & \textbf{0.6530}  & \textbf{0.6725}         \\
\hline
Variant A1  & 0.7051 &0.6502  & 0.6707           \\
Variant A2 & 0.7023 &0.6469 & 0.6674 \\
Variant A3 & 0.7030 & 0.6482 & 0.6681 \\
Variant A4 & 0.7067 &0.6510 & 0.6699 \\
\hline
Variant B1 & 0.7015 &0.6461 & 0.6651 \\
Variant B2 & 0.7043 &0.6473 & 0.6676 \\
Variant B3 & 0.7069 &0.6511 & 0.6708 \\
Variant B4 & 0.7062 &0.6517 & 0.6704 \\
\hline 
\end{tabular}
\label{tab:ablation}
\end{table}

\subsubsection{Evaluation Metrics}
In our offline experiments, we use the widely adopted area under the ROC curve (AUC) metric to evaluate model performance, where a higher AUC indicates better results. Additionally, following \cite{relaimpr1, din}, we employ the relative improvement (RelImpr) metric to compare performance gains between models. For the online A/B tests, we use CTR, cost per mille (CPM), and inference time as evaluation metrics.

\subsubsection{Implementation Details}
All models are implemented using TensorFlow. For fairness, all models are configured to have parameters of the same order of magnitude. In the overall performance experiments, the embedding size is set to 32 for all models, and the MLP in the prediction layer is configured as [32, 1] with Relu as the activation function. For our model and the models in Group \uppercase\expandafter{\romannumeral3}, the encoder consists of 2 layers and uses 8 attention heads.

Due to the varying lengths of user behavior sequences across different datasets, the sequence length is set to 1024 for all models on the industry and Alibaba datasets, and to 50 for the Ele.me dataset. All models are trained on NVIDIA A100-80G GPUs. Each model is trained for one epoch using the Adam optimizer.

\subsection{Overall Performance}
Table~\ref{tab:performance_comparison} presents the performance results of the competitors and SparseCTR across the three datasets. Each value is the average of the results obtained from repeating the same experiment three times. We also provide the standard deviation of AUCs of SparseCTR. It is worth noting that, in CTR prediction tasks, an AUC improvement at the 0.001 level is considered significant~\cite{twin, deepcross}. Besides, we conduct a $t$-test on the AUCs of SparseCTR and each baseline using a significance level of 0.05, and the fact that all $p$-values are less than 0.05 demonstrates that the performance differences between SparseCTR and the baselines are statistically significant.

Group \uppercase\expandafter{\romannumeral2} outperforms Group \uppercase\expandafter{\romannumeral1}, indicating that models designed for long-term behaviors attain better results on long sequences. Compared to Groups I and II, Group \uppercase\expandafter{\romannumeral3} achieves better results, confirming the effectiveness of self-attention mechanisms in modeling complex user behaviors. However, due to their high computational complexity, such models are challenging to deploy in real-world online systems.

SparseCTR outperforms all the baselines, indicating that our model not only significantly reduces computational complexity compared to the models in Group \uppercase\expandafter{\romannumeral3}, but also delivers superior AUCs. This advantage is primarily attributed to the EvoAttention and RelTemporal mechanisms, which effectively capture the sequential and periodic relationships among user behaviors.

\subsection{Ablation Study}
We construct two groups of variants of SparseCTR to evaluate the impact of each component in EvoAttention and RelTemporal on AUC. In the first group, variant A1 replaces the entire EvoAttention with standard full self-attention, while variants A2, A3, and A4 remove the global, transition, and local attention branches from EvoAttention, respectively. In the second group, variant B1 removes the complete RelTemporal, and variants B2, B3, and B4 exclude the relative time, hour, and weekend encodings, respectively. 

The results are shown in Table~\ref{tab:ablation}. All variant models perform worse than SparseCTR. For EvoAttention, both global and transition attention have a notable impact on performance. Removing local attention (variant A4) results in a small performance drop, presumably because transition attention can partially compensate for the absence of local attention. From the performance of variants in the second group, we find that RelTemporal indeed makes a substantial contribution to model performance, particularly through the relative time encoding. However, removing the relative hour or weekend encodings, both of which are used to capture periodic relationships among user behaviors, has only a limited effect on overall performance.

%Transition attention can partially compensate for the absence of local attention, so removing local attention (variant A4) results in a smaller performance drop. Regarding RelTemporal, it makes a substantial contribution to model performance, particularly through the relative time encoding. The relative hour and weekend encodings both capture periodic patterns, so removing either one has only a limited effect on overall performance.
\begin{figure}[t]
     \centering
         \includegraphics[width=0.95\linewidth]{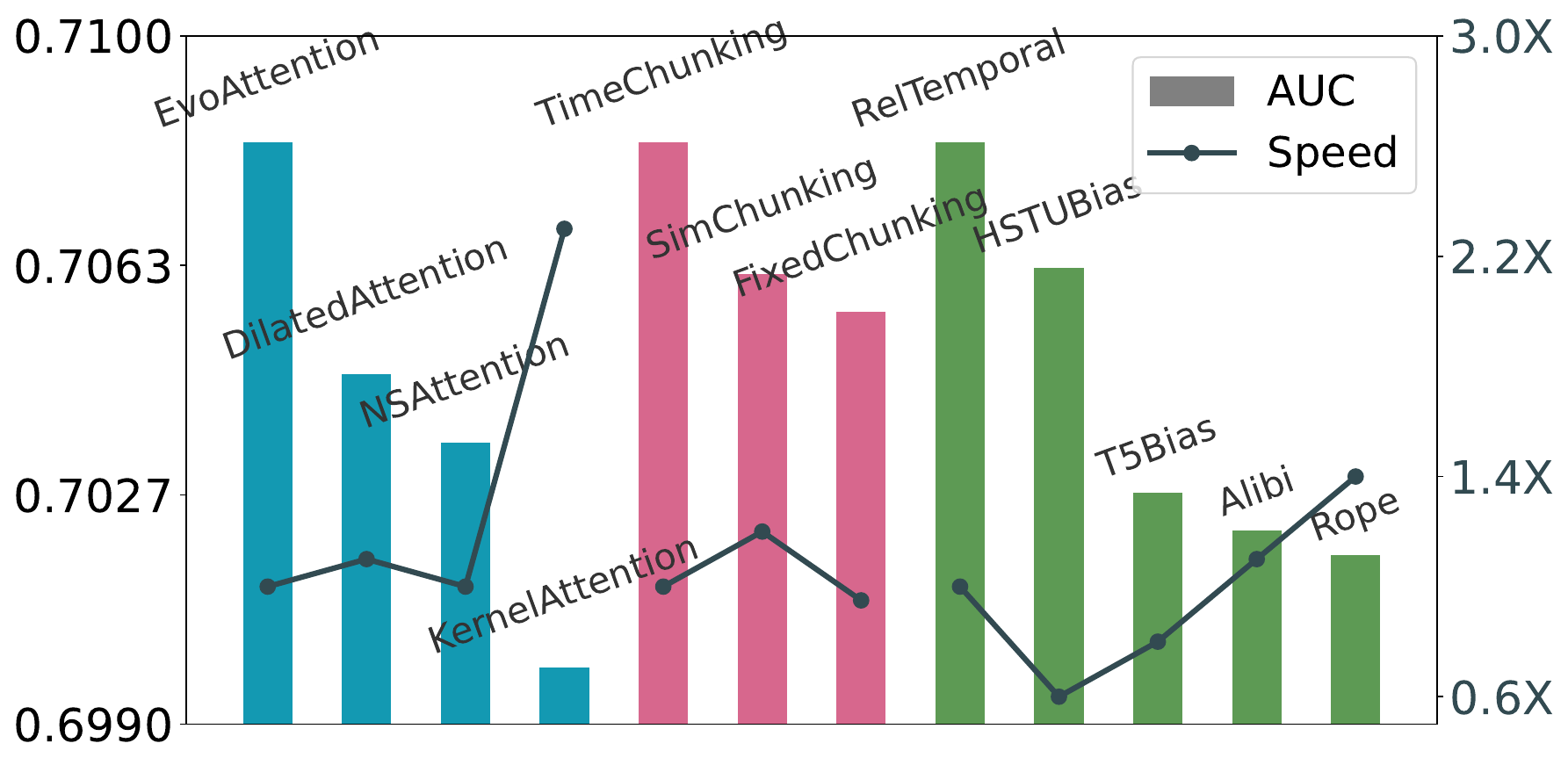}
         \caption{Comparison with existing sparse self-attention, chunking, and relative encoding methods.}
         \label{fig:compare}
\end{figure}

\subsection{Different Alternatives Comparison}
\subsubsection{Comparison of Different Sparse Self-attention Methods}
We compare EvoAttention with three representative sparse self-attention mechanisms, that is, 
the dilated attention (DilatedAttention)~\cite{dilated_sparse1}, the native sparse attention (NSAttention)~\cite{nsa}, 
%the dilated sparse method (DilatedAttention)~\cite{dilated_sparse1}, the fixed chunk method (NSAttention)~\cite{nsa}, 
and the kernelized attention method (KernelAttention)~\cite{Linformer}. Clustering attention~\cite{reformer} is excluded as it does not support causal modeling. To ensure fairness, we tune the hyperparameters so that EvoAttention, DilatedAttention, and NSAttention achieve comparable speeds. We record AUCs and speeds on the industrial dataset, where the speed is calculated based on the inference time of EvoAttention.

The results in the left section of Figure~\ref{fig:compare} show that EvoAttention outperforms both the DilatedAttention and NSAttention, as it accounts for user behavior distribution specific to recommender systems, making it more effective to model long-term user behaviors. Although KernelAttention achieves the highest efficiency, its AUC is significantly lower, as it is a simplified version of standard full self-attention and cannot incorporate relative time encoding.

\subsubsection{Comparison of Different Chunking Methods}
We compare personalized TimeChunking with the similarity-aware chunking (SimChunking), and fixed-length chunking (FixedChunking)~\cite{nsa} methods. The SimChunking method is similar to TimeChunking, except it uses embedding similarity as the criterion for chunking instead of time intervals. We also record AUCs and speeds on the industrial dataset, where the speed is calculated based on the inference time of TimeChunking.

As shown in the middle section of Figure~\ref{fig:compare}, TimeChunking achieves optimal performance with minimal computational overhead. This effectiveness is attributed to the personalized and temporal characteristics of user behaviors, allowing the TimeChunking to compress consecutive behaviors efficiently.

\subsubsection{Comparison of Different Relative Encoding Methods}
We compare RelTemporal to four relative information encoding methods, i.e., the attention bias approaches used in HSTU~\cite{hstu} and T5~\cite{t5}, as well as Alibi~\cite{alibi} and RoPE~\cite{rope}. 
%We record AUCs and speeds on the industrial dataset, where the speed is calculated based on the inference time of RelTemporal.

The experimental results are shown in the right section of Figure~\ref{fig:compare}. The results indicate that T5Bias, Alibi, and RoPE, each of which only considers relative positional relationships between behaviors, thereby show low AUCs. In contrast, HSTUBias, which incorporates relative time information, significantly improves performance but greatly reduces efficiency. RelTemporal not only achieves the highest AUC, but also maintains relatively high efficiency.

% \begin{table}
%     \centering
%         \caption{Configurations of different EvoAttention variants.}
%     \begin{tabular}{c|ccc}
%         \hline
%         Configuration & $|P|$ & $m$ & $w$ \\
%         \hline
%         Config 1 & 8 & 1 & 2\\
%         Config 2 & 16 & 2 & 5\\
%         Config 3 & 16 & 3 & 10\\
%         Config 4 & 32 & 4 & 20\\
%         \hline
%     \end{tabular}
%     \label{tab:evoattention}
% \end{table}

% \begin{table}
%     \centering
%     \caption{EvoAttention configurations with varying sparsity.}
%     \begin{tabular}{c|ccc}
%         \hline
%         Sparsity& $|P|$ & $m$ & $w$ \\
%         \hline
%         sparsity-1 & 8 & 1 & 2\\
%         sparsity-2 & 16 & 2 & 5\\
%         sparsity-3 & 16 & 3 & 10\\
%         sparsity-4 & 32 & 4 & 20\\
%         \hline
%     \end{tabular}
%     \label{tab:evoattention}
% \end{table}

\begin{figure}[t]
     \centering
         \includegraphics[width=0.8\linewidth]{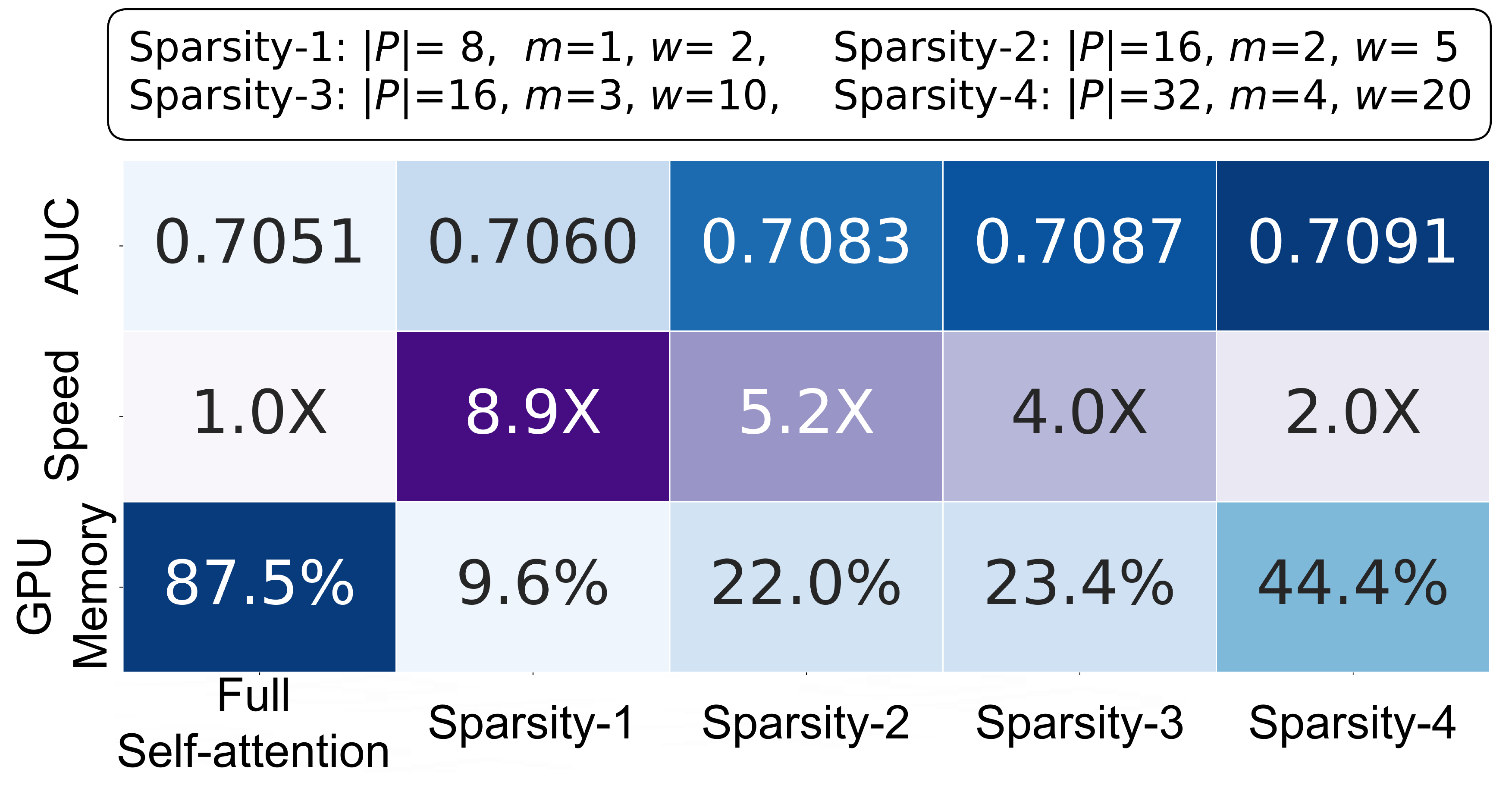}
         \caption{Performance of EvoAttention under different configurations in terms of AUC, speed, and GPU memory.}
         \label{fig:parameter}
\end{figure}
\subsection{Sparsity of Attention Analysis}
We tune the hyperparameters to obtain different configurations of EvoAttention with varying sparsity, as shown in the top line of Figure~\ref{fig:parameter}, where $|P|$ denotes the number of chunks, $m$ denotes the number of transition behaviors per chunk, and $w$ denotes the number of local behaviors. Then we run SparseCTR equipped with EvoAttention of varying sparsity and standard full self-attention on the industrial dataset, taking sequences of length 1024 as input, and record their AUCs, speeds, and GPU memory consumption.
%In this section, we tune the hyperparameters to obtain different configurations of EvoAttention with varying sparsity, and compare their performance to standard full self-attention on sequences of length 1024 with the same batch size. We evaluate AUC, speed, and GPU memory consumption. The specific configurations are presented in Table~\ref{tab:evoattention}, where $|P|$ denotes the number of chunks, $m$ the number of transition behaviors per chunk, and $w$ the number of local behaviors.

From the results of Figure~\ref{fig:parameter}, we find that all configurations of EvoAttention achieve higher AUC than standard full self-attention. This can be attributed to the fact that user behavior sequences often contain noise, and directly attending to all behaviors may lead to suboptimal results. More importantly, EvoAttention achieves up to an 8.9$\times$ speedup while using less GPU memory. This paves the way for large-scale deployment of long-sequence user modeling models in real-world systems.

\subsection{Scaling Law Analysis}
\begin{figure}[t]
     \centering
         \includegraphics[width=0.9\linewidth]{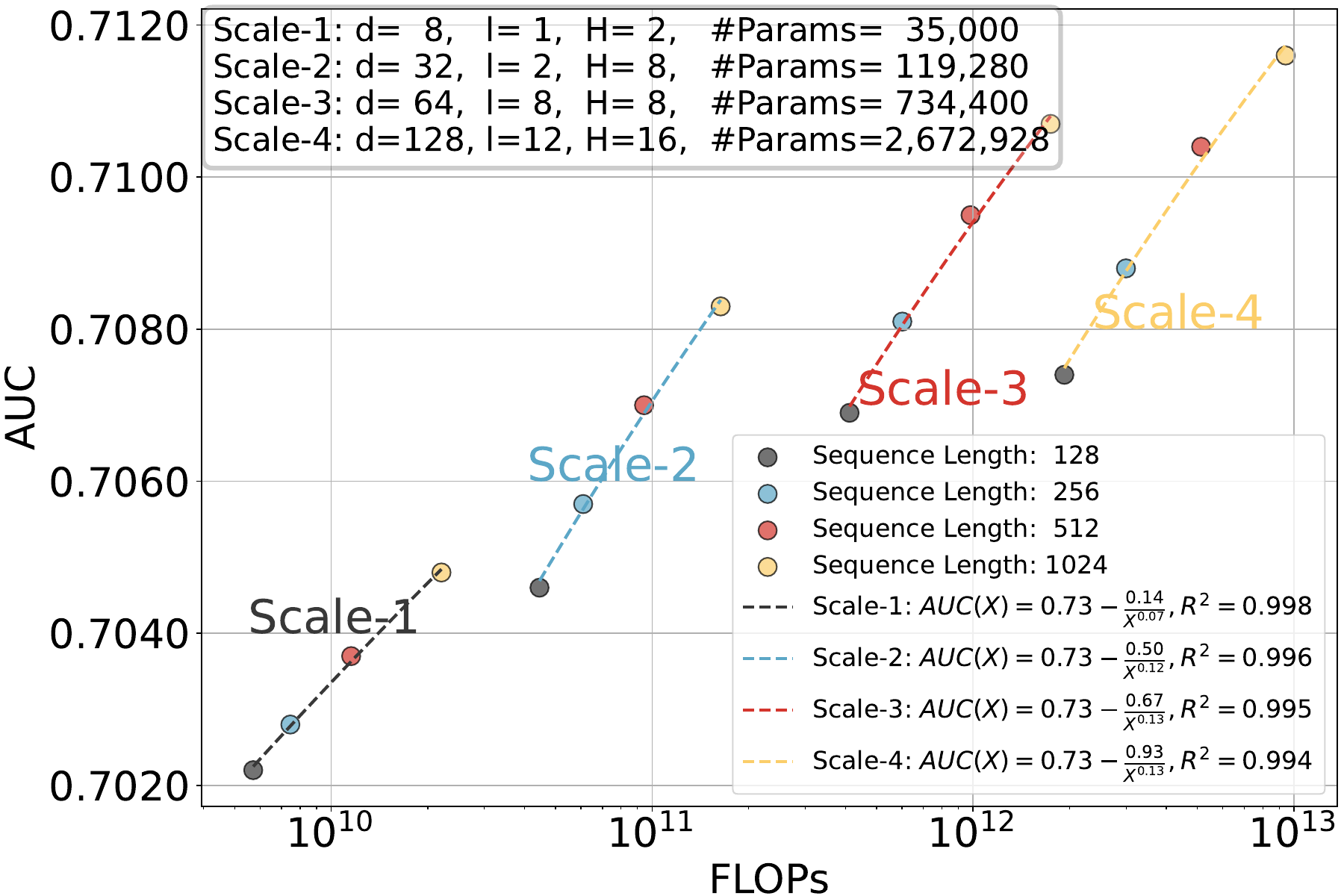}
         \caption{Scaling laws on the industrial dataset: AUC under different FLOPs, with FLOPs determined by sequence length and model scale (\#Params is the number of non-embedding parameters).}
         \label{fig:auc_flops}
\end{figure}

We conduct experiments on the industrial dataset to verify whether SparseCTR exhibits scaling laws. Specifically, we construct four models with different scales (with detailed configurations provided in the legend of Figure~\ref{fig:auc_flops}), each modeling user behavior sequences of varying lengths ([128, 256, 512, 1024]), so that our experiments span three orders of magnitude in FLOPs. Since AUC is a more meaningful metric in our scenario, we follow~\cite{scaling_law_funcion, SRT, SUAN} and 
fit the AUCs to a power-law function $AUC(X)=E-{A}/{X^\alpha}$, 
%fit the AUCs to a power-law function as follows:
%\begin{equation}
%    AUC(X)=E-{A}/{X^\alpha}
%\end{equation}
where $X$ denotes the FLOPs, and $E$ and $A$ are coefficients to be fitted. Here, $E$ can be interpreted as the potential upper bound of AUC values on these training samples.  

In the results shown in Figure~\ref{fig:auc_flops}, we
%The results are shown in Figure~\ref{fig:auc_flops}. We 
use the least squares method to fit the AUCs to a nonlinear curve, and the coefficients of determination ($R^2$) are close to 1, indicating a good fit. This demonstrates that SparseCTR preserves the key property that AUC increases as computational cost increases, highlighting the practical value of our model for real-world deployment. Furthermore, the fitted power-law functions provide insights into model scalability strategies; for example, simultaneously increasing both sequence length and model scale yields greater efficiency gains than increasing either one alone.
% \begin{table}
% \centering
% \caption{Configurations of different model scales.}
% \setlength\tabcolsep{4.2pt}
% \begin{tabular}{c|ccc|c}
% \hline
% \multirow{2}{*}{Model Scale} & \multirow{2}{*}{$d$} & \multirow{2}{*}{$l$} & \multirow{2}{*}{$H$} & \multirow{2}{*}{\shortstack{\#Non-embedding \\ Parameters}} \\
% & & & & \\  % This creates the empty line for multirow
% \hline
% scale-1  & 8  & 1  & 2  & 35,000       \\
% scale-2  & 32  & 2  & 8  & 119,280      \\
% scale-3  & 64  & 8  & 8  & 734,400       \\
% scale-4  & 128 & 12  & 16  & 2,672,928  \\
% \hline 
% \end{tabular}
% \label{tab:scale}
% %\vspace{-10pt}
% \end{table}

\subsection{Online A/B Test}
We conducted an online A/B test by deploying our SparseCTR model in a live production environment for 7 days, directing 1\% of the total traffic to it. 
%with 1\% traffic. 
The baseline, serving the list advertising scenario, is a well-established model that incorporates DIN, SIM, SUAN, and a standard full-attention component based on HSTU ( with sequence length 128), and other components, and has undergone several rounds of optimization. In the A/B test, we enhanced the baseline by replacing the online model's full-attention component with SparseCTR (with sequence length 1024). As a result, CTR improved by 1.72\% and CPM increased by 1.41\%, while the inference time maintains 40 ms.
%We conducted an online A/B test by deploying our SparseCTR model (with model scale 2 and sequence length 1024) in the live production environment for 7 days. The baseline, which serves the list advertising scenario, is a well-established model incorporating DIN, SIM, and several rounds of optimization. In the A/B test, we enhanced the baseline by integrating our model. As a result, CTR improved by XXX\% and CPM increased by XXX\%, while inference time only slightly increased from XXms to XXms.

\section{Conclusion}
In this paper, we propose the SparseCTR model for the CTR prediction task. The model learns user interests from diverse temporal
%spatiotemporal 
views within long-term user behaviors and presents a three-branch sparse attention mechanism, thereby avoiding the computational complexity that scales quadratically with the sequence length. Notably, SparseCTR exhibits the obvious scaling law phenomenon, enabling it to support longer sequences and larger parameters under the same computational resources, and achieving superior online performance.

%In this paper, we propose SparseCTR to alleviate the high computational complexity of modeling long-term user behaviors with large-scale models. SparseCTR incorporates a time-aware dynamic chunking method and integrates global, transition, and local attention mechanisms to efficiently capture user interests at multiple granularities. Furthermore, we introduce relative time, hour, and weekend information to explicitly model temporal and periodic patterns. Notably, our model maintains the obvious scaling law phenomenon even after sparsification, enabling SparseCTR to support longer sequences and larger models under the same computational resources, thereby achieving superior online performance.
\begin{acks} 
This work was supported by the National Natural Science Foundation of China under Grant No. 62072450 and Meituan.
\end{acks}

\bibliographystyle{ACM-Reference-Format}
\balance
\bibliography{sample-base}

\end{document}